\documentstyle[aps,12pt]{revtex}
\newcommand{\bra}[1]{\left\langle #1 \right|}         
\newcommand{\ket}[1]{\left| #1 \right\rangle}
\begin{document}
\title{Beyond the relativistic Hartree mean-field approximation:
energy dependent effective mass}
\author{D. Vretenar$^{1,2}$, T. Nik\v si\' c$^{1,2}$, and P. Ring$^{2}$}
\address{
1. Physics Department, Faculty of Science, 
University of Zagreb, 10000 Zagreb, Croatia\\
2. Physics Department, TU Munich, D-85748 Garching,
Germany \\
}
\date{\today}
\maketitle
\bigskip
\bigskip
\begin{abstract}
The standard relativistic mean-field model is extended 
by including dynamical effects that arise in the coupling
of single-nucleon motion to collective surface vibrations.
A phenomenological scheme, based on a linear ansatz for 
the energy dependence of the scalar and vector components
of the nucleon self-energy for states close to the Fermi
surface, allows a simultaneous description of binding 
energies, radii, deformations and single-nucleon spectra
in a self-consistent relativistic framework. The model
is applied to the spherical, doubly closed-shell nuclei
$^{132}$Sn and $^{208}$Pb.
\end{abstract}


%
%
\section {Introduction}

Relativistic models based on quantum hadrodynamics \cite{SW.97} provide 
a microscopic self-consistent description of 
the nuclear many-body problem.
Detailed properties of finite nuclei
along the $\beta$-stability line have been
described with nuclear structure models based on the 
relativistic mean-field (RMF) approximation~\cite{Rin.96}.
With the unified Dirac-Hartree-Bogoliubov description of 
mean-field and pairing correlations, the relativistic 
framework has been very successfully extended to studies
of exotic nuclei far from the valley of $\beta$- stability and to 
the physics of the drip 
lines~\cite{PVL.97,LVP.98,Lal.99,VLR.99,LVR.99,LVR.01}.

In this framework the single-nucleon dynamics 
is described by classical equations of motion
which are derived self-consistently from a fully relativistic Lagrangian.
Standard relativistic nuclear structure models are based on the
static approximation, i.e. the nucleon self-energy is real, local and
energy independent. Consequently, these models describe correctly the 
ground-state properties and the sequence of single-particle levels
in finite nuclei, but not the level density around the Fermi surface. 

In a non-relativistic mean-field approximation, the total effective
mass $m^*$ of a nucleon in a nucleus characterizes the energy 
dependence of an effective local potential that is equivalent 
to the, generally nonlocal and frequency dependent, microscopic
nuclear potential~\cite{MBB.85,MS.96}. The total effective mass 
is a product of the $k-mass$ which characterizes the non-locality, 
i.e. momentum dependence of the mass operator, and the $E-mass$ 
which describes the explicit energy dependence of the mass operator. 
The coupling of single-nucleon motion to 
collective vibrations and the resulting enhancement of the effective
mass around the Fermi surface has been extensively studied in 
the framework of non-relativistic Hartree-Fock models. For a review,
see Refs.~\cite{MBB.85} and \cite{MS.96}. 

In the context of the 
present analysis, two non-relativistic microscopic descriptions of
the energy dependent effective mass are of particular interest. 
The analyses of the neutron-$^{208}$Pb~\cite{JHM.87},
proton-$^{208}$Pb~\cite{MS.89},
and neutron-$^{40}$Ca~\cite{JM.88} mean fields were based on 
dispersion relations that connect the real and imaginary parts
of the optical-model potential. The parameters of the 
complex mean field were determined from available experimental 
cross sections, and the resulting optical-model potential was
extrapolated towards negative energies. A very good agreement 
was found with the experimental single-particle energies of the 
valence particle and hole states. 
In Refs.~\cite{BVG.79,BVG.80,VGVT.83}
the effects of the collective modes on the single-particle
states and the effective mass in $^{208}$Pb were calculated
in a self-consistent microscopic particle-vibration coupling 
model. It was shown that, as a function of energy 
and radial coordinate, the effective mass is enhanced around
the Fermi level and on the surface of the nucleus. A similar
approach was used in a very recent calculation of the 
nucleon $E-mass$ for a medium-heavy deformed nucleus as a 
function of the rotational frequency along the yrast line~\cite{Don.00}.

In the relativistic framework, the concept of effective nucleon
mass in symmetric nuclear matter and finite nuclei was analyzed
by Jaminon and Mahaux in Refs.~\cite{JM.89} and \cite{JM.90}.
In addition to the $k-mass$ and $E-mass$, a third effective mass,
the "Lorentz mass" appears in the relativistic approach. 
It results from different Lorentz transformation properties of
the scalar and vector potentials. 
In Ref.~\cite{JM.90} a quantitative analysis of the 
energy dependence of the effective mass was performed in the 
framework of the relativistic Brueckner-Hartree-Fock approximation
to the mean field in symmetric nuclear matter. The calculation was
based on dispersion relations that connect the imaginary 
to the real part of the Lorentz components of the mean field. 
Although in finite 
nuclei the mechanism which leads to the energy dependence of 
the effective mass is different, i.e. it results from the 
coupling of single-nucleon motion to the collective modes, 
a very useful result of Ref.~\cite{JM.90} is that the energy 
dependence of the total dispersive contribution to the 
real part of the mean field is almost linear in the region
$E_F - 10 < E < E_F + 10$ MeV around the Fermi energy $E_F$.

In the present work we propose a 
phenomenological scheme to include the effect
of coupling of single-nucleon motion to surface vibrations, i.e.
the energy dependence of the nucleon self-energy, in self-consistent
relativistic mean field calculations. Rather than calculating 
the first-order correction to the RMF single-nucleon energy spectra,
a linear energy dependence of the scalar and vector potentials 
is explicitly included in the Dirac equation. The resulting 
generalized eigenvalue problem can be solved exactly, and this
approach gives the possibility to reproduce, in a self-consistent
calculation, both the total binding energy and the density of 
single-nucleon states around the Fermi surface. The principal
advantage over perturbation calculations is, however, that the
present approach can be easily extended to constrained calculations
in deformed nuclei. The main purpose will be the study of shape 
coexistence phenomena in chains of isotopes far from the valley
of $\beta$-stability. 
%
%
%
\section {The relativistic mean-field model with energy 
dependent potentials}

In the framework of the relativistic mean field approximation
nucleons are described as point particles that
move independently in mean fields
which originate from the nucleon-nucleon interaction.
The theory is fully Lorentz invariant.
Conditions of causality and Lorentz invariance impose that the
interaction is mediated by the
exchange of point-like effective mesons, which couple to the nucleons
at local vertices. The single-nucleon dynamics is described by the
Dirac equation
\begin{equation}
\label{statDirac}
\left\{-i\mbox{\boldmath $\alpha$}
\cdot\mbox{\boldmath $\nabla$}
+\beta(m+g_\sigma \sigma)
+g_\omega \omega^0+g_\rho\tau_3\rho^0_3
+e\frac{(1-\tau_3)}{2} A^0\right\}\psi_i=
\varepsilon_i\psi_i.
\end{equation}
$\sigma$, $\omega$, and
$\rho$ are the meson fields, and $A$ denotes the electromagnetic potential.
$g_\sigma$ $g_\omega$, and $g_\rho$ are the corresponding coupling
constants for the mesons to the nucleon.
The lowest order of the quantum field theory is the {\it
mean-field} approximation: the meson field operators are
replaced by their expectation values. The sources
of the meson fields are defined by the nucleon densities
and currents.  The ground state of a nucleus is described
by the stationary self-consistent solution of the coupled
system of the Dirac~(\ref{statDirac})and Klein-Gordon equations:
\begin{eqnarray}
\left[ -\Delta +m_{\sigma }^{2}\right] \,\sigma ({\bf r}) &=&-g_{\sigma
}\,\rho _{s}({\bf r})-g_{2}\,\sigma ^{2}({\bf r})-g_{3}\,\sigma ^{3}({\bf r})
\label{messig} \\
\left[ -\Delta +m_{\omega }^{2}\right] \,\omega ^{0}({\bf r}) &=&g_{\omega
}\,\rho _{v}({\bf r})  \label{mesome} \\
\left[ -\Delta +m_{\rho }^{2}\right] \,\rho ^{0}({\bf r}) &=&g_{\rho }\,\rho
_{3}({\bf r})  \label{mesrho} \\
-\Delta \,A^{0}({\bf r}) &=&e\,\rho _{p}({\bf r}),  \label{photon}
\end{eqnarray}
for the sigma meson, omega meson, rho meson and photon field, respectively.
Due to charge conservation, only the 3rd-component of the isovector 
\mbox{\boldmath $\vec \rho$}
meson contributes. The source terms in equations (\ref{messig}) to (\ref
{photon}) are sums of bilinear products of baryon amplitudes, and they
are calculated in the {\it no-sea} approximation, i.e. the Dirac sea
of negative energy states does not contribute to the nucleon densities
and currents. Due to time reversal invariance,
there are no currents in the static solution for an even-even
system, and therefore the spatial
vector components \mbox{\boldmath $\omega,~\rho_3$} and
${\bf  A}$ of the vector meson fields vanish.
The quartic potential
\begin{equation}
U(\sigma )~=~\frac{1}{2}m_{\sigma }^{2}\sigma ^{2}+\frac{1}{3}g_{2}\sigma
^{3}+\frac{1}{4}g_{3}\sigma ^{4}  \label{usigma}
\end{equation}
introduces an effective density dependence. The non-linear
self-interaction of the $\sigma$ field is essential for
a quantitative description of properties of finite nuclei.

The details of the calculated ground-state properties
depend on the choice of the effective Lagrangian. Several
effective interactions, i.e. parameter sets of the mean-field
Lagrangian have been derived that provide a satisfactory
description of nuclear properties along the $\beta$-stability line.
In particular, the parameter set NL3~\cite{LKR.97} has been 
adjusted to ground state properties
of a large number of spherical nuclei. Properties
calculated with the NL3 effective interaction are found to
be in very good agreement with experimental data for 
spherical and deformed nuclei
at and away from the line of $\beta$-stability. In particular,
NL3 has been used in most applications of the Relativistic 
Hartree-Bogoliubov model to the physics of drip-line 
nuclei~\cite{PVL.97,LVP.98,Lal.99,VLR.99,LVR.99,LVR.01}.

The effective potential that determines the ground state of a finite 
nucleus is essentially given by the sum of the scalar 
$\sigma$-potential (attractive) and the vector 
$\omega$-potential (repulsive).
Both potentials are of the order of several hundred MeV 
in the nuclear interior. The contributions of the isovector 
$\rho$-meson field and the electromagnetic interaction are, 
of course, much smaller. In the relativistic 
Hartree mean-field approximation the nucleon self-energy is real, local and
energy independent. It should be emphasized, however, 
that even in the Hartree approximation the equivalent Schr\" odinger 
potential is nonlocal, i.e. momentum dependent. This results 
from the momentum dependence of the scalar density, or equivalently, 
the momentum dependence of the Dirac mass in the non-relativistic 
reduction of the Dirac equation.

A phenomenological description of the effect of coupling between 
single-nucleon motion and collective modes can be obtained by 
assuming that the scalar and vector potentials depend linearly 
on energy in the vicinity of the Fermi surface
\begin{eqnarray}
V_\sigma(r,E) &=& V_\sigma(r) + \alpha (E - E_F)\nonumber \\ 
V_\omega(r,E) &=& V_\omega(r) + \alpha (E - E_F), 
\label{potE}
\end{eqnarray}
where $V_\sigma(r)$ and $V_\omega(r)$ denote the usual,
energy independent potentials.
For simplicity, we only consider the energy dependence of 
the two most important contributions to the effective
potential, i.e. the $\sigma$ and $\omega$ fields.
As we have already emphasized in the Introduction, in their 
analysis of the Fermi surface anomaly and depletion of the 
Fermi sea in the relativistic Brueckner-Hartree-Fock 
approximation~\cite{JM.90}, Jaminon and Mahaux have shown
that the total dispersive contribution to the real part 
of the mean field potential displays an almost linear 
energy dependence in a region of $\approx~\pm 10$ MeV around
the Fermi energy. A linear energy dependence of the real 
part of the mean field potential in the region
$E_F - 5 < E < E_F + 5$ was also found in the calculation of 
the nucleon effective mass in $^{208}$Pb in the framework of 
the microscopic particle-vibration coupling model
(see Fig. 1 of Ref.~\cite{VGVT.83}). If the effective potential
in the Dirac equation depends only linearly on energy, this 
defines a generalized eigenvalue problem
\begin{equation}
H_D \ket{\psi} = A E \ket{\psi} 
\label{gep}
\end{equation}
where $H_D$ is the energy independent Dirac hamiltonian 
(see Eq. (\ref{statDirac})), and
on the right-hand side $A\times E$ is the matrix which 
contains the linear energy dependence. Eq. (\ref{gep}) can 
be solved exactly, and in this way the effect of coupling to 
collective modes can be included in the self-consistent
calculation of the ground state of a nucleus.

In principle, there is no deeper reason for the $\sigma$ and $\omega$
potentials to have the same energy dependence. If they have, like 
in Eqs. (\ref{potE}), the total correction to the effective 
single-nucleon potential is $2\alpha (E - E_F)$, and there is 
no correction to the spin-orbit term of the effective potential,
which is given by the difference of the $\sigma$ and $\omega$
potentials~\cite{Rin.96}. Since in the present analysis we
are only interested in the effect 
on the density of states around the Fermi level,
we assume the same linear energy dependence for 
$V_\sigma$ and $V_\omega$. This means that, in principle, an 
additional parameter can be adjusted to obtain, if necessary,
a better agreement for the energy splittings of the 
spin-orbit partner levels in spherical or deformed nuclei.

The ground state of a spherical or deformed nucleus is obtained
from a self-consistent solution of the coupled system of equations:
the Dirac generalized eigenvalue
equation (\ref{gep}), and the Klein-Gordon equations (\ref{messig})-
(\ref{photon}). The equations are solved by expanding 
the nucleon spinors and the meson fields in a spherical or deformed
harmonic oscillator basis. In the present work we only consider 
spherical nuclei. Instead of calculating an energy dependent
correction to single nucleon spectra determined by some of the 
standard RMF parameter sets such as NL3, we include
the energy dependent potentials in the fitting procedure that is
used to construct an effective interaction. By adjusting the 
meson-nucleon coupling constants, the mass of the $\sigma$ meson,
the parameters of the $\sigma$ meson self-interaction terms, 
and the parameter $\alpha$ of the linear energy dependence 
in Eqs. (\ref{potE}), to nuclear matter and properties of 
finite spherical nuclei, we seek to obtain a simultaneous 
description of binding energies and of densities of single-nucleon
states in the valence shells.

In order to illustrate the method, in the present analysis we 
calculate the single-nucleon spectra of the doubly closed-shell
nuclei $^{208}$Pb and $^{132}$Sn. In this particular calculation 
the starting point is the NL3 parameter set. This effective 
interaction has been adjusted to ground-state properties of
a number of spherical nuclei and, of course, it reproduces 
the binding energies of $^{208}$Pb and $^{132}$Sn. The calculated
densities of single-nucleon states around the Fermi surface are,
however, too low as compared with the experimental spectra,
i.e. the effective mass is too low and energy independent.
By including the linear energy dependence (\ref{potE}) in the 
effective single-nucleon potential of the Dirac equation, the 
binding is obviously reduced and the parameter set of the 
effective interaction has to be readjusted. In principle, the
parameter of the energy dependence $\alpha$ in Eqs. (\ref{potE})
can be calculated from the imaginary part of the optical 
potential by using dispersion relations, or by coupling 
single-nucleon states to core vibrations calculated with
the relativistic random phase approximation. That would, of 
course, mean a different $\alpha$ for every nucleus. In the 
present self-consistent method $\alpha$ is an adjustable 
parameter, determined by a fit to ground state properties,
together with the parameter set of the effective Lagrangian.
In this work we only calculate $^{208}$Pb and $^{132}$Sn,
although of course a larger set of nuclei would be 
necessary to obtain an optimal set of parameters. For 
closed-shell nuclei without pairing, the Fermi energy is 
taken as the half-energy between the last occupied and
the first unoccupied single-nucleon orbit. The linear
energy correction to the effective potential is confined
to the window $E_F - 10 < E < E_F + 10$ MeV. In addition to the 
available binding energies and radii, the fitting procedure is 
constrained with the average energies of hole and particle
states in the last occupied and the first unoccupied major
shells, respectively.

In each iteration step, and for given single-nucleon 
angular momentum and parity, the generalized eigenvalue
problem (\ref{gep}) is solved in two steps. In the 
first step the single-nucleon 
energy independent Dirac hamiltonian $H_D$
(see Eq. (\ref{statDirac})) is diagonalized in the 
harmonic oscillator basis. From the resulting 
single-nucleon spectrum we determine the eigenvectors
$\psi_i$ with eigenvalues $\epsilon_i$ in the 
energy interval $\approx~\pm 10$ MeV around
the Fermi energy. In the next step the matrix of the 
generalized eigenvalue problem is constructed
\begin{equation}
H_{ij}~=~
\left\{
\begin{array}{cc}
\bra{\psi_i} H_D \ket{\psi_j} & \\
\bra{\psi_i} H_D + 2\alpha (E-E_f) \ket{\psi_j}
& {\rm for}~E_F - 10 < \epsilon_i, \epsilon_j < E_F + 10~{\rm MeV}
\end{array}
\right .
\end{equation}
For a solution in coordinate space, the equivalent of this
two-step procedure is the inclusion of the theta-function
$\Theta(10~{\rm MeV} - |E - E_F|)$ in front of the 
energy dependent correction in (\ref{potE}). Although 
we do not include an explicit radial dependence for 
the second term in (\ref{potE}), its contribution 
obviously vanishes at large distances, i.e. 
all matrix elements $H_{ij}$ vanish for large
$r$ because $\psi_i (\vec r) \rightarrow 0$.

The set of states which is the 
solution of the generalized eigenvalue problem 
is not orthogonal. This is not a big effect, however,  
because the energy dependent correction to the potential
is relatively small, and it affects
only a small number of states in the vicinity 
of the Fermi level. After the orthogonalization,
the new set of orthonormalized states is identified
with the single-nucleon spectrum. There is no significant
difference between the diagonal matrix elements of the hamiltonian 
in this basis, i.e. the "single-particle energies",
and the original eigenvalues.
We have verified that the same value for the 
ground-state energy is obtained when this quantity 
is calculated as the sum of kinetic and potential 
energies, or as the sum of the "single-nucleon 
energies" and the energies of the meson fields.

The resulting single-neutron spectra in $^{208}$Pb and $^{132}$Sn
are displayed in Figs. \ref{figA} and \ref{figB}, respectively.
they are compared with the relativistic Hartree mean-field 
results calculated with the NL3 interaction, and with 
experimental single-neutron spectra~\cite{1,2,15,18,21,22}. 
In comparison with the original NL3 interaction, we note
a considerable improvement of the calculated spectra. 
The increase of the density of states around the Fermi surface
results from the energy dependence of the effective mass. 
The spectra obtained with the full potential are in good
agreement with the experimental single-neutron levels.
The energy dependent correction, however, seldom
changes the ordering of states, and therefore we still find
inverted doublets, as for example $2f_{5/2}$-$3p_{3/2}$ and
$1i_{11/2}$-$2g_{9/2}$ in $^{208}$Pb. In Table \ref{tabA}
we compare the NL3 effective interaction with the new 
set of parameters which is obtained by including the 
energy dependent correction to the single-nucleon potential
in the fitting procedure. We note that the values of the 
parameters change very little as compared with the original
NL3 parameterization, with the exception of a somewhat
more pronounced decrease of the rho-meson coupling. The adjusted
value of the coefficient of the linear energy dependence in 
Eqs. (\ref{potE}) is $2 \alpha = -0.288$. This value can 
be compared with the one
calculated in the self-consistent microscopic particle-vibration
coupling model of Refs.~~\cite{BVG.79,BVG.80,VGVT.83}. 
For proton states $E_F - 5 < E < E_F + 5$ MeV
in $^{208}$Pb (see Fig. 1 of Ref.~\cite{VGVT.83}),
the coefficient of linear energy dependence of the real 
part of the mass operator is 
$\approx -0.35$. In addition to
single-nucleon spectra, we also compare the experimental
and calculated binding energies: for $^{208}$Pb
the experimental value is -1636.47 MeV, the value 
calculated with NL3 is -1639.54 MeV, 
and the full, energy dependent effective
potential gives -1636.80 MeV. For $^{132}$Sn, the binding energy calculated
with NL3 is -1105.44 MeV, the value obtained with the new parameter
set is -1101.31 MeV, and the experimental binding energy is 
-1102.90 MeV. 

The single-neutron spectrum of $^{208}$Pb can be compared with 
those obtained with the self-consistent 
microscopic particle-vibration coupling
model (see Fig. 2 of Ref.~\cite{BVG.80}), and with the phenomenological
dispersion relation analysis of Ref.~\cite{JHM.87} (see Fig. 9 of that
article). In comparison with the experimental spectrum, the results
of the present calculation are somewhat better than those 
obtained with the microscopic particle-vibration coupling, 
but not as good as the spectrum which results from dispersion
relations between the real and imaginary parts of the optical-model
potential. This is, of course, not very surprising. In the 
analysis of Ref.\cite{BVG.80} the mass operator 
is the sum of a Hartree-Fock term and an energy dependent term which 
arises from the coupling to the microscopically calculated RPA
vibrations. A number of approximations have to be made in such 
a microscopic calculation, which will necessarily affect the 
quality of the final single-nucleon spectrum. On the other
hand, there is no adjustable parameter like $alpha$ in
the microscopic calculation. In the phenomenological
analysis of Ref.~\cite{JHM.87} the total
dispersive contribution to the effective potential is written 
as a sum of volume and surface components. A relatively large
number of parameters was used to adjust these two contributions 
separately. It was shown that about 40\% of the total
correction results from volume dispersive effects and about 60\%
from the surface-peaked correction. The comparison with the
single-neutron spectra of Refs.~\cite{BVG.80} and \cite{JHM.87},
as well as with the experimental data, shows that our linear 
ansatz (\ref{potE}) presents a very good approximation of
the dynamical effects which arise from the coupling of 
single-nucleon and collective degrees of freedom.

In Tables~\ref{tabB} and \ref{tabC} we list both the neutron 
and the proton levels of $^{208}$Pb and $^{132}$Sn, respectively.
For the last occupied and first unoccupied major shells, we compare
the results calculated with the standard
relativistic Hartree mean-field approximation (NL3 interaction),
the  energy spectrum
calculated with the full, energy dependent model potential, 
and the experimental energies. In addition, we compare the 
average energies of hole (particle) states 
\begin{equation}
<E> = {{\sum_{nlj} (2j+1) E_{nlj}}\over {\sum_{nlj}  (2j+1) }},
\end{equation}
where the sum runs over occupied (unoccupied) states within a
major shell.
Obviously the inclusion of linear energy dependence in the nucleon 
self-energy considerably improves the calculated single-nucleon spectra,
while at the same time it enables the self-consistent calculation of global
ground state properties, such as the binding energy and radii.
%
%
\section {Conclusions}

In applications of standard relativistic mean-field models
to the description of ground state properties of spherical
and deformed nuclei, the nucleon self-energy is real, local and
energy independent. This leads to the well known problem of 
low effective mass, i.e. low density of single-nucleon 
states close to the Fermi surface. In the non-relativistic
framework, this problem has been considered in first-order
perturbation, either by explicitly coupling the single-nucleon
states to collective RPA modes, or by using dispersion relations
that connect the real and imaginary parts of the optical-model
potential. In the present work we have studied a phenomenological
scheme which allows the inclusion of the dynamical effects 
of coupling of single-nucleon motion to surface vibrations
in self-consistent relativistic mean field calculations.
The scheme is based on a linear ansatz for the energy 
dependence of the scalar and vector components of the 
nucleon self-energy for states close to the Fermi surface.
This defines a generalized Dirac 
eigenvalue problem, which can be solved self-consistently
together with the Klein-Gordon equations for the meson fields.
Thus, rather than calculating the first-order correction
to the single-nucleon spectra, the assumption of linear
energy dependence for the mass operator enables a
self-consistent
calculation of both the global ground state properties 
(masses, radii) and the single-nucleon levels around the
Fermi surface. The model that we have studied might be 
especially useful in studies of shape-coexistence phenomena
in nuclei far from the valley of $\beta$-stability. 
Because of the simple linear energy dependence of 
the effective single-nucleon potential, the method 
can be easily extended to constrained calculations in
deformed nuclei. 

In the present work the method has been tested on the
spherical, doubly closed-shell nuclei $^{132}$Sn and $^{208}$Pb.
The ground-state properties of these nuclei are well
described by the standard NL3 effective interaction
of the relativistic mean-field model. With the inclusion of linear
energy dependent terms in the scalar and vector
components of the nucleon self-energy for states close to the Fermi
surface, the parameter set of the effective Lagrangian has to be
readjusted in order to reproduce
both the global ground state properties and 
the densities of single-nucleon states. 
The resulting single-nucleon spectra have been compared 
with experimental data, as well as with previous
non-relativistic first-order perturbation calculations 
for $^{208}$Pb. It has been shown that the 
energy dependent effective mass considerably 
improves the calculated single-nucleon spectra.
By allowing the simultaneous description of binding
energies, radii and single-nucleon spectra, the 
self-consistent method studied in this work
presents a natural phenomenological extension 
of the relativistic mean-field model. Work is in progress 
on the derivation of the energy dependent effective mass by 
coupling single-nucleon states to surface vibrations calculated
in the relativistic random phase approximation, and on the
extension of the model to deformed nuclei.

\bigskip

{\bf ACKNOWLEDGMENTS}

This work has been supported in part by the 
Bundesministerium f\"{u}r Bildung und Forschung 
under the project 06 TM 979.  D.V. and T.N. would like to acknowledge 
the support from the Alexander von 
Humboldt - Stiftung.

\newpage
\begin{figure}
\caption{Neutron single-particle states in $^{208}$Pb.
In the first column on the left the relativistic Hartree mean-field 
results calculated with the NL3 effective interaction are displayed,
the levels calculated with the energy dependent effective
potential are shown in the second column, and the right-hand 
column gives the experimental spectrum~\protect\cite{21,22}.}
\label{figA}
\end{figure}

\begin{figure}
\caption{Same as in Fig.~\protect\ref{figA}, 
but for the neutron single-particle states in $^{132}$Sn.
The experimental data are from Refs.~\protect\cite{1,2,15,18}.}
\label{figB}
\end{figure}

\begin{table}
\caption{The parameter set of the NL3 effective 
interaction~\protect\cite{LKR.97} (center column), and the 
new interaction (right column) which results from the inclusion
of the linear energy dependence in the effective single-nucleon 
potential.}

\begin{center}
\begin{tabular}{ccc}
                &     NL3        &      NEW     \\ \hline
\hline
{m}		&     {939.0 MeV}   &     {939.0 MeV} \\ 
{$m_{\sigma}$}  &     {508.1941 MeV}   &     {508.8500 MeV} \\ 
{$m_{\omega}$}  &     {782.5010 MeV}   &     {782.5550 MeV} \\ 
{$m_{\rho}$}    &     {763.0 MeV}   &     {763.0 MeV} \\ 
{$g_{\sigma}$}  &     {10.2169}        &     {10.2200}      \\ 
{$g_{\omega}$}  &     {12.8675}	       &     {12.8730}      \\ 
{$g_{\rho}$}    &     {4.4744}	       &     {4.4000}       \\ 
{$g_{2}$}	&     {-10.4307} fm$^{-1}$    
 	&     {-10.2153} fm$^{-1}$    \\ 
{$g_{3}$}	&     {-28.8851}       &     {-29.0960}     \\ 
{$2 \alpha$}      &                      &     {-0.288}      \\
\end {tabular}
\label{tabA}
\end{center}
\end{table}

\begin {table}[h]
\caption{Neutron (left panel) and proton (right panel) single-particle
energies $E_{nlj}$ in $^{208}$Pb. For each panel, the left-hand column 
specifies the radial, orbital and total angular momentum
quantum numbers, the column labeled NL3 contains results calculated
with the standard 
relativistic Hartree mean-field approximation, the energy spectrum
calculated with the full, energy dependent, model potential is displayed
in the third column. Theoretical spectra are compared to the experimental
energies shown in the column labeled EXP. With boldface letters are listed
the average energies for particle and hole states.}

\begin{center}
\begin{tabular}{c c c c c c c c }
\multicolumn{4}{c}{\sc neutron states}  &
\multicolumn{4}{c}{\sc proton states} \\    
\hline
{$nlj$}	         & {\sc nl3}      & {\sc full}     & {\sc exp}  
& {$nlj$}	 & {\sc nl3}      & {\sc full}     & {\sc exp}  \\ \hline
\hline
{$4s_{1/2}$}     & {-0.36}   & {-1.46}   & {-1.90} 
& {$3p_{1/2}$}     & { 2.58}   & { 0.60}   & {-0.17} \\ \hline
{$3d_{3/2}$}     & {-0.02}   & {-1.19}   & {-1.40} 
& {$3p_{3/2}$}     & { 1.83}   & { 0.00}   & {-0.68} \\ \hline
{$3d_{5/2}$}     & {-0.63}   & {-1.72}   & {-2.37}  
& {$2f_{5/2}$}     & { 0.55}   & {-0.92}   & {-0.98} \\ \hline
{$2g_{7/2}$}     & {-0.57}   & {-1.60}   & {-1.44}  
& {$2f_{7/2}$}     & {-1.44}   & {-2.55}   & {-2.90}  \\ \hline
{$2g_{9/2}$}     & {-2.50}   & {-3.17}   & {-3.94}  
& {$1h_{9/2}$}     & {-4.60}   & {-4.97}   & {-3.80}  \\ \hline
{$1i_{11/2}$}    & {-2.97}   & {-3.29}   & {-3.16}  
& {$1i_{13/2}$}    & {-1.03}   & {-2.37}   & {-2.19}  \\ \hline
{$1j_{15/2}$}    & {-0.48}   & {-1.60}   & {-2.51}  
& {\boldmath $E_{p}$} &  {\bf -1.28}  & {\bf -2.45}   &  {\bf -1.79}  \\ \hline
{\boldmath $E_{p}$} & {\bf -1.33}   & {\bf -2.20}   & {\bf -2.63}  
&                   &           &           &            \\ \hline
\hline
{$3p_{1/2}$}     & {-7.67}   & {-7.34}   & {-7.37}  
& {$3s_{1/2}$}     & {-8.15}   & {-7.95}   & {-8.01}   \\ \hline
{$3p_{3/2}$}     & {-8.41}   & {-7.94}   & {-8.26}  
& {$2d_{3/2}$}     & {-9.25}   & {-8.79}   & {-8.36}  \\ \hline
{$2f_{5/2}$}     & {-9.09}   & {-8.38}   & {-7.94}  
& {$2d_{5/2}$}     & {-10.88}  & {-10.16}  & {-9.68}   \\ \hline
{$2f_{7/2}$}     & {-11.11}  & {-10.05}  & {-9.71}  
& {$1g_{7/2}$}    & {-15.05}  & {-13.55}   & {-11.48} \\ \hline
{$1h_{9/2}$}     & {-13.39}  & {-11.78}  & {-10.78} 
& {$1h_{11/2}$}    & {-10.21}  & {-9.82}   & {-9.35} \\ \hline
{$1i_{13/2}$}    & {-9.60}   & {-8.97}   & {-9.00}  
& {\boldmath $E_{h}$} &  {\bf -11.30} & {\bf -10.57}  &  {\bf -9.38} \\ \hline
{\boldmath $E_{h}$} & {\bf -10.47}  & {\bf -9.55}   & {\bf -9.25}  
&                 &           &           &          \\ 
\end {tabular}
\end{center}
\label{tabB}
\end{table}

\begin {table}[h]
\caption{Same as in Table~\protect\ref{tabB}, 
but for the neutron and proton single-particle states in $^{132}$Sn.}

\begin {center}
\begin{tabular}{c c c c  c c c c }
\multicolumn{4}{c}{\sc neutron states}  &
\multicolumn{4}{c}{\sc proton states} \\    
\hline
{$nlj$}	         & {\sc nl3}      & {\sc full}     & {\sc exp}  
& {$nlj$}	 & {\sc nl3}      & {\sc full}     & {\sc exp}  \\ \hline
\hline
{$2f_{5/2}$}     & {0.08}    & {-0.96}   & {-0.58} 
& {$3s_{1/2}$}   & {-4.22}   & {-6.27}   & {-6.83}  \\ \hline
{$3p_{1/2}$}     & {-0.26}   & {-1.25}   & {-0.92}  
& {$1h_{11/2}$}    & {-5.33}   & {-7.33}   & {-6.84} \\ \hline
{$1h_{9/2}$}     & {-0.46}   & {-1.21}   & {-1.02}  
& {$2d_{3/2}$}     & {-5.29}   & {-7.08}   & {-7.19} \\ \hline
{$3p_{3/2}$}     & {-0.57}   & {-1.50}   & {-1.73}  
& {$2d_{5/2}$}     & {-7.05}   & {-8.56}   & {-8.67}  \\ \hline
{$2f_{7/2}$}     & {-1.33}   & {-2.12}   & {-2.58}  
& {$1g_{7/2}$}     & {-9.95}   & {-10.77}  & {-9.63} \\ \hline
{\boldmath $E_{p}$} & {\bf -0.59}   & {\bf -1.44}   & {\bf -1.03}  
& {\boldmath $E_{p}$} & {\bf -6.73}   & {\bf -8.32}   & {\bf -7.92}  \\ \hline
\hline
{$2d_{3/2}$}     & {-8.76}   & {-8.03}   & {-7.31}  
& {$1g_{9/2}$}     & {-16.12}  & {-16.06}  & {-15.71} \\ \hline
{$1h_{11/2}$}    & {-7.65}   & {-7.33}   & {-7.55}  
& {$2p_{1/2}$}     & {-17.11}  & {-16.65}  & {-16.07} \\ \hline
{$3s_{1/2}$}     & {-8.33}   & {-7.71}   & {-7.64}  
& {\boldmath $E_{h}$} & {\bf -16.28}  & {\bf -16.15}  & {\bf -15.77} \\ \hline
{$2d_{5/2}$}     & {-10.48}  & {-9.47}   & {-8.96}  
&                   &           &           &          \\ \hline
{$1g_{7/2}$}     & {-12.31}  & {-10.84}  & {-9.74}  
&                   &           &           &          \\ \hline
{\boldmath $E_{h}$} & {\bf -10.65}  & {\bf -9.55}   & {\bf -8.81}  
&                   &           &           &           \\ 
\end {tabular}
\label{tabC}
\end{center}
\end{table}
\end{document}